\documentstyle[eqsecnum,aps,prd,epsf]{revtex}
\def\sqr#1#2{{\vcenter{\hrule height.#2pt
      \hbox{\vrule width.#2pt height#1pt \kern#1pt
          \vrule width.#2pt}
      \hrule height.#2pt}}}

\def\edth{{\rlap{$\partial$}\raise0.3em\hbox{$-$}}}

\def\gtrless{{\hbox{\raisebox{0.6ex}{$\,>$}}\kern-1.8ex
                   \hbox{\raisebox{-0.6ex}{$<\,$}}}}

\begin{document}
\thispagestyle{empty}
\vspace*{-10mm}
\epsfxsize=2cm
\leftline{\epsfbox{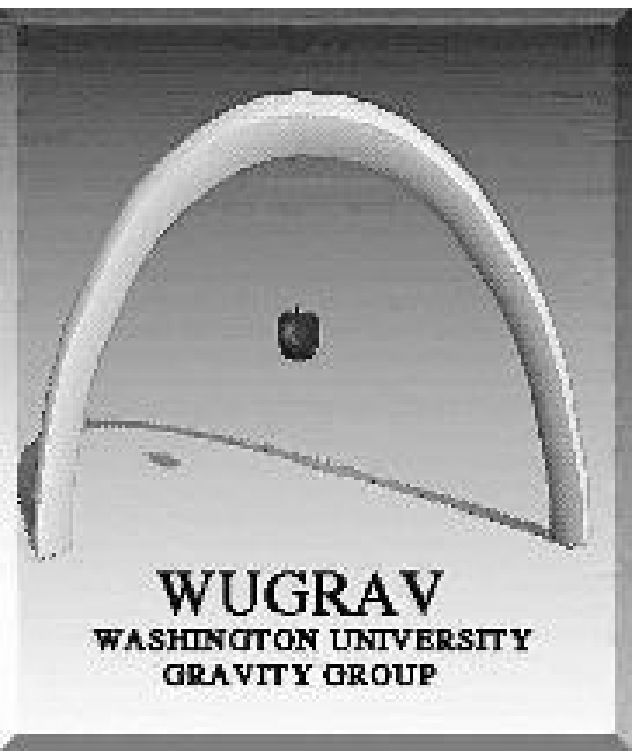}}
\vspace{5mm}

{\baselineskip-4pt
\font\yitp=cmmib10 scaled\magstep2
\font\elevenmib=cmmib10 scaled\magstep1  \skewchar\elevenmib='177
\leftline{\baselineskip20pt
\vbox to0pt
   { \hbox{{\yitp Washington \hspace{1.5mm} University \hspace{1.5mm}
St}.\hspace{1.5mm}{\yitp Louis} }
     {\large\sl\hbox{{WUGRAV}} }\vss}}

\vspace*{-20mm}
\begin{flushright}
UTAP-386\\
OUTAP-157
\end{flushright}
\vspace*{20mm}

\begin{center}
{\Large \bf Covariant Self-force Regularization of a Particle \\
Orbiting a Schwarzschild Black Hole}

\vspace{3mm}

{\Large \it - Mode Decomposition Regularization -}

\bigskip

Yasushi Mino$^{1}$, Hiroyuki Nakano$^{2}$,
and Misao Sasaki$^{2}$

\smallskip

$^1${\em Department of Physics, Washington University,\\
Campus Box 1105, One Brookings Dr., St. Louis, MO 63130-4899, USA}\\
\smallskip
$^1${\em Department of Physics, Faculty of Science, University of Tokyo,\\
Bunkyo-ku, Tokyo, 113-0033 Japan}\\
\smallskip
$^1${\em Theoretical Astrophysics, California Institute of Technology,
Pasadena CA 91125, USA}\\
\smallskip
$^2${\em Department of Earth and Space Science,~Graduate School of
  Science,~Osaka University,\\ Toyonaka, Osaka 560-0043, Japan
}\\
\smallskip

\medskip

\today

\bigskip

{\bf abstract}

\end{center}

{\small
Covariant structure of the self-force of a particle in a general
curved background has been made clear in the cases of scalar\cite{Quinn},
electromagnetic\cite{DeWittBrehme}, and gravitational charges
\cite{reaction,QuinnWald}. Namely, what we need is the
part of the self-field that is non-vanishing off and within
the past light-cone of particle's location, the so-called tail.
The radiation reaction force in the absence of external fields
is entirely contained in the tail.
In this paper, we develop mathematical tools for the regularization
and propose a practical method to calculate the
self-force of a particle orbiting a Schwarzschild black hole.
}


\section{Introduction}\label{sec:intro}

For a particle carrying a scalar, electromagnetic or
gravitational charge, the field configuration of
the corresponding type varies in time
as it moves around a black hole.
To the lowest order in the charge, the particle motion
follows a geodesic in the black hole background
in the absence of external force fields.
However, a part of the time-varying
field becomes radiation near the future null infinity or
future horizon and carries the energy-momentum away
from the system, and a part of it is scattered by the
background curvature and comes back
to the location of the particle. Hence the
motion of the particle is affected in the next order.
The force exerted by the back-scattered self-field
is called the local reaction force or simply the self-force.
To establish a calculational strategy of this force is
our ultimate goal.

It is noted that we may consider the local reaction force to
consist of the two parts: The part that describes
the loss of the energy-momentum of the particle and
the other part that only contributes to the shift of
conserved quantities. In the case of geodesic motion in
the Schwarzschild background, this division is unambiguous
because the geodesic motion is completely determined by
the two conserved quantities; the energy and the $z$-component
of the angular momentum (a geodesic can be assumed to be
on the equatorial plane without loss of generality).
In the case of the Kerr background, however, it seems unclear
if the two parts can be identified uniquely because of the
presence of the conserved quatity called the Carter constant which
has no explicit relation to the energy-momentum of the system.

When we attempt to calculate the reaction force on a point charge
(particle), we encounter the divergence of the force.
Hence, it is necessary to extract out the physically meaningful
finite part of the force. Since the force is a vector by definition
with respect to a background space-time,
and any vector depends on the choice of coordinates in a covariant manner,
the finite reaction force should be given covariantly.

The covariant structure of the reaction force was investigated
in the scalar case in \cite{Quinn},
in the electromagnetic case in \cite{DeWittBrehme},
and in the gravitational case in \cite{reaction,QuinnWald}.
In these investigations, the divergent part of the force was found
to be described solely with the local geometrical quantities,
whereas the finite part that contributes to the equation of motion
was found to be given by the tail part which
is due to the curvature scattering of the self-field.

Since the tail part depends non-locally on the geometry of the
background spacetime, it is almost impossible to calculate it
directly. However, for a certain class of spacetimes such as
Schwarzschild/Kerr geometries, there is a way to calculate
the full field generated by a point charge.
Considering a field point slightly off the particle trajectory,
it is then possible to obtain the tail part by subtracting
the locally given divergent part from the full field.
Thus denoting the field by ${}_s\phi$ for the scalar
($s=0$), electromagnetic ($s=1$) or gravitational ($s=2$) case,
with its spacetime indices suppressed,
the reaction force is schematically given by
\begin{eqnarray}
F_\alpha(\tau_0) &=&
\lim_{x\rightarrow z(\tau_0)}F_\alpha[{}_{s}{\phi}^{\rm tail}](x)
\,, \label{eq:scheme} \\
F_\alpha[{}_{s}{\phi}^{\rm tail}](x) &=&
F_\alpha[{}_{s}{\phi}^{\rm full}](x)-F_\alpha[{}_{s}{\phi}^{\rm dir}](x)
\,, \quad (x \not = z(\tau)) \,,
\end{eqnarray}
where $z$ is the orbit of the particle with the proper time $\tau$,
and $\tau_0$ is the proper
time at the orbital point at which we calculate the force. The symbol
${}_{s}{\phi}^{\rm tail}$ stands for the tail field induced by the particle
which is regular in the coincidence limit $x \to z(\tau)$,
${}_{s}{\phi}^{\rm full}$ for the full field,
and ${}_{s}{\phi}^{\rm dir}$ for the direct part as defined
in Refs.\cite{Quinn,DeWittBrehme,reaction,QuinnWald}.
Both ${}_{s}{\phi}^{\rm full}$ and ${}_{s}{\phi}^{\rm dir}$ diverge
in the coincidence limit $x\to z(\tau)$.
$F_\alpha[...]$ is a tensor operator on the field, and is defined as
\begin{eqnarray}
F_\alpha[{}_{s}{\phi}] &=& \displaystyle\cases{
q P_\alpha{}^\beta \nabla_\beta \phi & ($s=0$), \cr
e P_\alpha{}^\beta (\phi_{\gamma;\beta}-\phi_{\beta;\gamma})V^\gamma
& ($s=1$), \cr
-m P_\alpha{}^\beta \left(\phi_{\beta\gamma;\delta}
-{1\over 2}g_{\beta\gamma}\phi^\epsilon{}_{\epsilon;\delta}
-{1\over 2}\phi_{\gamma\delta;\beta}
+{1\over 4}g_{\gamma\delta}\phi^\epsilon{}_{\epsilon;\beta}\right)
V^\gamma V^\delta & ($s=2$),}
\label{eq:force}
\end{eqnarray}
where $P_\alpha{}^\beta = \delta_\alpha{}^\beta +V_\alpha V^\beta$
is the projection tensor with
$V^\alpha$ being an appropriate extension of
the four velocity $v^\alpha(\tau_0)$ off the orbital point.

In practice, it is a non-trivial task to perform
the subtraction of the direct part, which we call
the `{\it Subtraction Problem}'.
In this paper we propose a method to carry out this
subtraction procedure covariantly.

It should be noted, however, that solving the subtraction
problem is not enough when one deals with the
gravitational case. In the scalar or electromagnetic case,
the reaction force is a gauge-invariant notion. In contrast,
the reaction force in the gravitational case does depend
on the gauge choice.
Therefore, one has to fix the gauge appropriately and evaluate
the full metric perturbation and its direct part in the same gauge
before calculating the force. We call this the `{\it Gauge Problem}',
which seems to be a very difficult problem to solve.
We do not discuss the possible solution of the gauge problem in this paper,
but leave it for future work.
Instead, we only calculate the direct part of the linear gravitational force
under the harmonic gauge condition.

In this paper, as a first step, we consider the case
that the background is approximated by the Schwarzschild blackhole
and use the Boyer-Lindquist coordinates,
\begin{eqnarray}
ds^2 &=&
-\left(1-{2M\over r}\right)dt^2+\left(1-{2M\over r}\right)^{-1} dr^2
+r^2\left(d\theta^2+\sin^2\theta d\phi^2\right) \,.
\end{eqnarray}
We use the notation that $x=\{t,\,r,\,\theta,\,\phi\}$ stands for
a field point,
and $z(\tau_0)=z_0=\{t_0,\,r_0,\,\theta,\,\phi\}$ for an orbital point.

The paper is organized as follows. In Section \ref{sec:scheme},
we describe basics of the issue.
In Section \ref{sec:mode}, we describe our strategy of regularization,
which we call the `mode decomposition regularization'.
In Section \ref{sec:decomposition}, focusing on the scalar case,
we present our method for
the mode-decomposition of the direct part of the self-force.
In doing so, we obtain a useful mathematical formula,
Eq.~(\ref{eq:xipdec}), for the harmonic decomposition.
In Section \ref{sec:comp}, summing over the azimuthal modes $m$,
we compare our result with Barack and Ori\cite{BMONS,BOnew}.
We find complete agreement between the two.
In Section \ref{sec:concl}, we discuss the pros and cons
and conclude the paper by pointing out some future issues.
Technical details as well as the mode decomposition regularization
of the electromagnetic and gravitational cases are differed in Appendices.

\section{Basics} \label{sec:scheme}

There are two important issues in the regularization.
One is how to implement a regularization method in Eq.~(\ref{eq:scheme}).
Both ${}_{s}{\phi}^{\rm full}(x)$ and ${}_{s}{\phi}^{\rm dir}(x)$
diverge in the coincidence limit $x \to z_0$.
Such divergent quantities are difficult to treat in actual calculations,
particularly in numerical computations.
We discuss this issue in Subsection \ref{subsec:expansion}.
The other is how to evaluate the direct part of the field,
which we discuss in Subsection \ref{subsec:direct}.

\subsection{Infinite series expansion} \label{subsec:expansion}

We consider the case of a particle in geodesic orbit
on the Schwarzschild spacetime.
Even in the Newtonian limit, the integration of the orbit
involves an elliptic function.
Thus, numerical computaions will be necessary at some stage
of deriving the self-force.
However, we have divergence to be regularized which is difficult
to treat numerically.
The idea to overcome this difficulty is to replace
the divergence by an infinite series, each term of which is finite.

Suppose we have a unique decomposition method
applicable to $F_\alpha[{}_{s}{\phi}^{\rm full}](x)$,
$F_\alpha[{}_{s}{\phi}^{\rm dir}](x)$ and
$F_\alpha[{}_{s}{\phi}^{\rm tail}](x)$ as
\begin{eqnarray}
F_\alpha[{}_{s}{\phi}^{\rm full}](x) &=&
\sum_A F^A_\alpha[{}_{s}{\phi}^{\rm full}](x) \,,
\\
F_\alpha[{}_{s}{\phi}^{\rm dir}](x) &=&
\sum_A F^A_\alpha[{}_{s}{\phi}^{\rm dir}](x) \,,
\\
F_\alpha[{}_{s}{\phi}^{\rm tail}](x) &=&
\sum_A F^A_\alpha[{}_{s}{\phi}^{\rm tail}](x) \,.
\end{eqnarray}
Because of the uniqueness of the decomposition, we have
\begin{eqnarray}
F^A_\alpha[{}_{s}{\phi}^{\rm tail}](x) &=&
F^A_\alpha[{}_{s}{\phi}^{\rm full}](x)-F^A_\alpha[{}_{s}{\phi}^{\rm dir}](x)
\,.
\end{eqnarray}
At this stage, we assume that each term of the infinite series is finite,
then it is possible take the coincidence limit $x \to z_0$
since $F^A_\alpha[{}_{s}{\phi}^{\rm tail}](x)$ is guaranteed to be finite.
Therefore we have
\begin{eqnarray}
F_\alpha(\tau_0) &=& \sum_A F^A_\alpha[{}_{s}{\phi}^{\rm tail}](z_0) \,.
\label{eq:scheme1}
\end{eqnarray}
This approach itself does not justify a numerical method
in the regularization calculation.
However, because of the convergence in the infinite sum (\ref{eq:scheme1}),
we can expect that the sum of a finite number of terms in (\ref{eq:scheme1})
gives us an approximated value of the self-force $F_\alpha(\tau_0)$.

However, there is a very delicate problem in this approach.
The exact decomposition calculation usually needs
the global analytic structure of the field
so that we can uniquely define each term in the infinite series.
On the other hand, the regularization scheme is derived just
by the local analysis of the field
\cite{Quinn,DeWittBrehme,reaction,QuinnWald}.
Thus the direct part is defined
only in the local neighborhood of the particle,
and we have an ambiguity in the definition of the direct part.
Because of this ambiguity,
each term in the infinite series expansion is no more unique
but depends on a global extension of the direct part we adopt.
Nevertheless, the final result of the self-force should be unique.

In this paper, we present a decomposition method based on the
spherical harmonic series expansion.
Although we have no explicit proof for the uniqueness of the
resulting regularization counter terms for the self-force,
the fact that our result completely agrees
with that of Barack and Ori\cite{BMONS,BOnew} strongly supports
the validity of our method.

\subsection{Direct Part of the Scalar Field} \label{subsec:direct}

The derivation of the direct part ${}_{s}{\phi}^{\rm dir}(x)$ is
one of the main issues in the regularization calculation.
The direct part of the scalar field is obtained
by integrating the direct part of the retarded Green function
with the source charge. Here we focus on the scalar case.
The electromagnetic and gravitational cases are treated
in the same manner, details of which are given in Appendix~\ref{app:EMG}.

The direct part of the retarded Green function $G^{\rm dir}$
is given in a covariant manner as
\begin{eqnarray}
G^{\rm dir}(x,x') &=& -{1\over 4\pi}\, \theta[\Sigma(x),x']
\sqrt{\Delta(x,x')}\delta\bigl(\sigma(x,x')\bigr)
\,, \label{eq:direct-g}
\end{eqnarray}
where $\sigma(x,x')$ is the bi-scalar of half the squared geodesic distance,
$\Delta(x,x')$ is the generalized van Vleck-Morette determinant,
$\Sigma(x)$ is an arbitrary spacelike hypersurface containing $x$,
and $\theta[\Sigma(x),x']=1-\theta[x',\Sigma(x)]$ is equal to $1$
when $x'$ lies in the past of $\Sigma(x)$
and vanishes when $x'$ lies in the future.
We summarize the basic properties of the bi-scalars
$\sigma(x,x')$ and $\Delta(x,x')$ in Appendix \ref{app:bi-tensor}.

The physical meaning of the direct part is understood by the factor
$\theta[\Sigma(x),x']\delta\bigl(\sigma(x,x')\bigr)$ in
Eq.~(\ref{eq:direct-g}).
Since $\sigma(x,x')$ describes the geodesic distance between $x$ and $x'$,
the direct part of the Green function becomes non-zero
only when $x'$ lies on the past lightcone of $x$.
Hence the direct part describes the effect
of the waves propagated directly from $x'$ to $x$
without scattered by the background curvature.

For the actual evaluation of the direct part,
several methods have been proposed.
In Ref.~\cite{Burko,Barack}, the direct part of the field is
calculated by picking up
a limiting contribution in the full Green function from the light cone as
\begin{eqnarray}
\phi^{\rm dir}(x)= \lim_{\epsilon\to+0}
\int^\infty_{\tau_{\rm ret}(x)-\epsilon}d\tau \,
G^{\rm full}(x,z(\tau))S(\tau)
\,,
\end{eqnarray}
where $G^{\rm full}$ is the retarded Green function,
$S(\tau)$ is the scalar charge density,
and $\tau_{\rm ret}(x)$ is the retarded time defined
by the past light cone condition of the field point $x$ as
\begin{eqnarray}
\theta[\Sigma(x),z(\tau_{\rm ret})]
\delta\bigl(\sigma(x,z(\tau_{\rm ret}))\bigr)=0 \,.
\end{eqnarray}
A number of works have been made along this approach\cite{Burko,Barack}.
However, the caluculation seems rather cumbersome
when we apply this method to a general orbit.

In Ref.~\cite{MinoNakano},
the direct part was evaluated using the local bi-tensor expansion
technique.
Using the bi-tensor,
the direct part is expanded around the particle location as
\begin{eqnarray}
\phi^{\rm dir}(x) &=&
q\left[{1 \over \sigma_{;\alpha}(x,z(\tau_{\rm ret}))
v^\alpha(\tau_{\rm ret})}\right] +O(y^2) \,,
\label{eq:direct}
\end{eqnarray}
where the letters $\mu$, $\nu$, $\cdots$ are used for the indices
of the field point $x$,
$\alpha$, $\beta$, $\cdots$ for the indices of the orbital point $z$,
and $v^\alpha(\tau)$ is the orbital four velocity at $z(\tau)$.
The order of the local expansion is represented by
powers of $y$ which is linear to the coordinate difference between
the field point $x$ and the orbital point $z_0$.
Because the full force is quadratically divergent,
we must carry out the local bi-tensor expansion of the full field
up through $O(y)$.

By evaluating the local coordinate values of the relevant bi-tensors,
we obtain the local expansion of the full force in a given coordinate
system (see Appendix~\ref{app:bi-tensor}).
As described in Ref.~\cite{MinoNakano}, this may be done in a
systematic manner, and it is possible to obtain the explicit
form of the divergence for a general orbit.
However, the problem is how to decompose it
into an appropriate infinite series. This is done in
Section \ref{sec:decomposition}.

\section{Mode decomposition regularization} \label{sec:mode}

We call the regularization calculation using the spherical
harmonic expansion by the mode decomposition regularization.
In this section, we briefly describe the regularization procedure
in this approach.

The harmonic decomposition is defined
by the analytic structure of the field on the two-sphere.
However both the direct field and the full field have a divergence
on the sphere including the particle location,
the mode decomposition is ill-defined on that sphere.
Therefore, we perform the harmonic decomposition of the direct and full fields
on a sphere which does not include, but sufficiently close to the orbit.
The steps in the mode decomposition regularization are as follows.

\begin{list}{}{}
\item[1)]
We evaluate both the full field and the direct field at
\begin{eqnarray}
x &=& \{t,r,\theta,\phi\} \,,
\end{eqnarray}
where we do not take the coincidence limit of either $t$ or $r$

\item[2)]
We decompose the full force and direct force
into infinte harmonic series as
\begin{eqnarray}
F_\alpha[{}_{s}{\phi}^{\rm full}](x) &=&
\sum_{\ell m}F_\alpha^{\ell m}[{}_{s}{\phi}^{\rm full}](x) \,, \\
F_\alpha[{}_{s}{\phi}^{\rm dir}](x) &=&
\sum_{\ell m}F_\alpha^{\ell m}[{}_{s}{\phi}^{\rm dir}](x) \,,
\end{eqnarray}
where $F_\alpha[{}_{s}{\phi}^{\rm full/dir}](x)$ are expanded
in terms of the spherical harmonics
$Y_{\ell m}(\theta,\phi)$
with the coefficients dependent on $t$ and $r$.\footnote{Rigorously
speaking, the angular components of the force is expanded as
$F_A=C_{\ell m}Y_{A,\ell m}$ where $Y_{A,\ell m}$ ($A=\theta,\phi$)
are the vector spherical harmonics.
We also note that
$F_\alpha[{}_{s}{\phi}_{\ell m}](x)$ and
$F_\alpha^{\ell m}[{}_{s}{\phi}](x)$ are different
since the tensorial property of the operator $F_\alpha[...]$
depends on the spin $s$ of the field ${}_s\phi$.
}
For the direct part, the harmonic expansion is done by extending
the locally defined direct force over to the whole two-sphere
in a way that correctly reproduces the divergent behavior around
the orbital point $z_0$ up to the finite term.

\item[3)]
We subtract the direct part from the full part in each $\ell$, $m$ mode
to obtain
\begin{eqnarray}
F_\alpha^{\ell m}[{}_{s}{\phi}^{\rm tail}] &=&
(F_\alpha^{\ell m}[{}_{s}{\phi}^{\rm full}]
-F_\alpha^{\ell m}[{}_{s}{\phi}^{\rm dir}]) \,.
\end{eqnarray}
The we take the coincidence limit $x\to z_0$.
Here we note that one can exchange the order of the procedure,
i.e., first take the coincidence limit and then
subtract, provided the mode coefficients are finite in the coincidence
limit.

\item[4)]
Finally, by taking the sum over the modes, we obtain
the self-force as
\begin{eqnarray}
F_\alpha(\tau_0) &=&
\sum_{\ell m}F_\alpha^{\ell m}[{}_{s}{\phi}^{\rm tail}](z_0) \,.
\end{eqnarray}
\end{list}

It should be noted that
because of the divergence of the full force and direct force
along a timelike orbit,
the mode coefficients of the full force and the direct force
are not uniquely defined when we take the coincidence limit in 3).
However, the tail force is regular
along the orbit\cite{Quinn,DeWittBrehme,reaction,QuinnWald},
and it is uniquely defined.
Therefore we expect the non-uniqueness of the direct force
does not cause a problem as long as the cincidence limit
is taken consistentlly for both the full force
and the direct force.

\section{Decomposition of the direct part} \label{sec:decomposition}

The advantage of using (\ref{eq:direct}) is that
we have a systematic method for evaluating the direct part,
which we describe in Subsection \ref{subsec:local}.
In Subsection \ref{subsec:dir_har},
we describe our method for the harmonic decomposition of the direct part.

\subsection{Local coordinate expansion} \label{subsec:local}

Though we have the covariant form of the
local bi-tensor expansion of the direct part,
it is not useful for the derivation of the infinite series expansion of it
until we evaluate it in a specific coordinate system.
Here we discuss the method to evaluate the bi-tensors
in a general regular coordinate system.

Before we consider the local expansion in a given coordinate system,
we calculate the derivative of (\ref{eq:direct}),
and derive the direct part of the force with the local bi-tensor expansion
using the equal-time condition,\footnote{Actually,
the equal-time condition is not essential
for the evaluation of the direct part.
However, one can see the dependence on the spin of the field
in a transparent manner under the equal-time condition
as we describe in Appendix~\ref{app:EMG}.}
\begin{eqnarray}
0 &=& \left[{d\over d\tau}\sigma(x,z(\tau))\right]_{\tau=\tau_{\rm eq}(x)}
\,. \label{eq:equal}
\end{eqnarray}
We define the extension of the four-velocity off the orbit by
\begin{eqnarray}
V^\alpha(x) :=
{\bar g}_{\alpha{\bar \alpha}}(x,z_{\rm eq})v^{\bar\alpha}_{\rm eq}
\,,
\label{eq:ppexV}
\end{eqnarray}
where ${\bar g}_{\alpha{\bar \alpha}}$ is the
parallel displacement bi-vector, $z_{\rm eq}=z(\tau_{\rm eq}(x))$,
and $v^{\bar\alpha}_{\rm eq}
=dz^{\bar\alpha}/d\tau|_{\tau=\tau_{\rm eq}(x)}$.
Using the formulas in Ref.~\cite{DeWittBrehme,reaction}, we have
\begin{eqnarray}
F_\alpha[\phi^{\rm dir}](x) &=&
q {\bar g}_\alpha{}^{\bar \alpha}(x,z_{\rm eq}){1\over \epsilon^3\kappa}
\left\{\sigma_{;\bar\alpha}(x,z_{\rm eq})
+{1 \over 3}\epsilon^2R_{\bar\alpha\bar\beta\bar\gamma\bar\delta}(z_{\rm eq})
v^{\bar\beta}_{\rm eq}\sigma^{;\bar\gamma}(x,z_{\rm eq})
v^{\bar\delta}_{\rm eq}\right\}+O(y)
\,, \label{eq:dir_force} \\
\epsilon &=& \sqrt{2\sigma(x,z_{\rm eq})}
\,, \\
\kappa &=& \sqrt{-\sigma_{\bar\alpha\bar\beta}(x,z_{\rm eq})
v^{\bar\alpha}_{\rm eq}v^{\bar\beta}_{\rm eq}}
\nonumber \\
&=& 1+{1\over 6}R_{\bar\alpha\bar\beta\bar\gamma\bar\delta}(z_{\rm eq})
v^{\bar\alpha}_{\rm eq}\sigma^{;\bar\beta}(x,z_{\rm eq})
v^{\bar\gamma}_{\rm eq}\sigma^{;\bar\delta}(x,z_{\rm eq}) +O(y^3)
\,.
\end{eqnarray}


The bi-tensors neccesary for the evaluation of
the direct force (\ref{eq:dir_force})
are $\sigma(x,\bar x)$ and ${\bar g}_{\alpha {\bar \alpha}}(x,\bar x)$,
which satisfy
\begin{eqnarray}
&& \sigma(x,{\bar x})
~=~ {1\over 2}g^{\alpha\beta}
\sigma_{;\alpha}(x,{\bar x})\sigma_{;\beta}(x,{\bar x})
~=~ {1\over 2}g^{{\bar \alpha}{\bar \beta}}
\sigma_{;{\bar \alpha}}(x,{\bar x})\sigma_{;{\bar \beta}}(x,{\bar x})
\,, \label{eq:bi_sigma} \\
&& \lim_{x\rightarrow{\bar x}}\sigma_{;\alpha}(x,{\bar x})
~=~ \lim_{x\rightarrow{\bar x}}\sigma_{;{\bar \alpha}}(x,{\bar x})
~=~ 0
\,, \\
&& {\bar g}_{\alpha{\bar \alpha};\beta}(x,{\bar x})
g^{\beta\gamma}(x)\sigma_{;\gamma}(x,{\bar x}) ~=~ 0 \,,\quad
{\bar g}_{\alpha{\bar \alpha};{\bar \beta}}(x,{\bar x})
g^{{\bar \beta}{\bar \gamma}}({\bar x})
\sigma_{;{\bar \gamma}}(x,{\bar x}) ~=~ 0
\,, \label{eq:bi_displacement} \\
&& \lim_{x\rightarrow{\bar x}}{\bar g}_\alpha{}^{\bar \alpha} ~=~
\delta_\alpha{}^{\bar \alpha} \,.
\end{eqnarray}
In addition, we need the generalized van Vleck-Morette determinant,
\begin{eqnarray}
\Delta(x,{\bar x}) &=&
\det\bigl(-{\bar g}^{\alpha{\bar\alpha}}(x,{\bar x})
\sigma_{;{\bar\alpha}\beta}(x,{\bar x})\bigr) \,.
\end{eqnarray}
We consider the local coordinate expansion of these bi-tensors
around the coincidence limit $x\rightarrow {\bar x}$,
assuming that we have no coordinate singularity at $\bar x$.

In the coincidence limit, the effect of the curvature is small
and we know the exact forms for half the geodesic distance bi-scalar
and the paralell displacement bi-vector
in the locally Cartesian coordinates as
\begin{eqnarray}
\sigma(x,{\bar x}) &=& {1\over 2}
\eta_{\alpha\beta}(x^\alpha-{\bar x}^\alpha)(x^\beta-{\bar x}^\beta)
+O(|x-{\bar x}|^3) \nonumber \,, \\
\bar g_{\alpha\bar\alpha}(x,{\bar x}) &=& \eta_{\alpha\bar\alpha}
+O(|x-{\bar x}|)  \,.
\end{eqnarray}
Therefore, in a general regular coordinate system,
$\sigma(x,{\bar x})$ and $\bar g_{\alpha\bar\alpha}(x,{\bar x})$
can be expanded as
\begin{eqnarray}
\sigma(x,{\bar x}) &=&
{1\over 2}g_{\alpha\beta}({\bar x})y^{\alpha\beta}
+\sum_{n=3,4,\cdots}{1\over n!}A_{\alpha^1\alpha^2\cdots\alpha^n}({\bar x})
y^{\alpha^1\alpha^2\cdots\alpha^n} \,, \label{eq:sigma-exp}
\\
{\bar g}_{\alpha{\bar\alpha}}(x,{\bar x}) &=&
g_{\alpha{\bar\alpha}}({\bar x})
+\sum_{n=1,2,\cdots}{1\over n!}
B_{\alpha{\bar\alpha}|\beta^1\beta^2\cdots\beta^n}({\bar x})
y^{\beta^1\beta^2\cdots\beta^n} \,, \label{eq:g-exp}
\end{eqnarray}
where
\begin{eqnarray}
y^{\alpha^1\alpha^2\cdots} = (x^{\alpha^1}-{\bar x}^{\alpha^1})
(x^{\alpha^2}-{\bar x}^{\alpha^2})\cdots.
\end{eqnarray}

To calculate the reaction force to a monopole particle,
it is enough to know the expansion coefficients
of $n=3,4$ of Eq.~(\ref{eq:sigma-exp}) and $n=1,2$ of
Eq.~(\ref{eq:g-exp}).\footnote{In the calculation of the direct force
given below, only Eq.~(\ref{eq:sigma-exp}) but Eq.~(\ref{eq:g-exp})
turns out to be necessary.
This is a result of our choice of the off-world line extension 
of the direct force, i.e., the parallel-propagation extension (\ref{eq:ppexV}).}
For a general metric, from (\ref{eq:bi_sigma}) and (\ref{eq:bi_displacement}),
we have
\begin{eqnarray}
A_{\alpha\beta\gamma} &=& {3\over 2}g_{(\alpha\beta,\gamma)} \,,
\label{eq:A3} \\
A_{\alpha\beta\gamma\delta} &=&
2g_{(\alpha\beta,\gamma\delta)}
-g_{\mu\nu}\Gamma^\mu_{(\alpha\beta}\Gamma^\nu_{\gamma\delta)} \,,
\label{eq:A4} \\
B_{\alpha\beta|\gamma} &=&
\Gamma_{\beta|\alpha\gamma} \,,
\label{eq:B3} \\
B_{\alpha\beta|\gamma\delta} &=&
{1\over 2}\left(\Gamma_{\beta|\alpha\gamma,\delta}
+\Gamma_{\beta|\alpha\delta,\gamma}
-g_{\mu\nu}\Gamma^\mu_{\alpha\gamma}\Gamma^\nu_{\beta\delta}
-g_{\mu\nu}\Gamma^\mu_{\alpha\delta}\Gamma^\nu_{\beta\gamma}
\right) \,. \label{eq:B4}
\end{eqnarray}
The explicit evaluation of these coefficients in the
Boyer-Lindquist coordinates is given in Appendix \ref{app:bi-tensor}.

The local expansion of the force (\ref{eq:dir_force})
on the Boyer-Lindquist coordinates is quite tedious, though systematic.
We implement this calculation using {\it Maple(R)}, an algebraic calculation
program, and have 100 pages output in the end.
However, most of the terms make a vanishing contribution
to the harmonic coefficients in the coincidence limit $x \to z_0$.
Below, we shall focus on the terms that are non-vanishing
in the coincidence limit.

\subsection{Harmonic decomposition of the direct part}
\label{subsec:dir_har}

Without loss of generality, we may assume that the particle
is located at $\theta_0=\pi/2$, $\phi_0=0$ at time $t_0$.
Since the full force is calculated in the form of the Fourier-harmonic
expansion and the Fourier modes are independent of the spherical
harmonics, we may take the field point to lie on the hypersurface $t=t_0$
in the full force. Hence we may take $t=t_0$ before we perform
the local coordinate expansion of the direct force.
That is, we consider the local coordinate expansion of the direct force
at a point $\{t_0,r,\theta,\phi\}$
near the particle localtion $\{t_0,r_0,\pi/2,0\}$.

The local expansion of the direct force on the
Boyer-Lindquist coordinates can be done in such a way that
it consists of terms of the form,
\begin{eqnarray}
{R^{n_1}\Theta^{n_2}\phi^{n_3}\over \xi^{2n_4+1}}\,,
\label{eq:ingredient}
\end{eqnarray}
where $n_1$, $n_2$, $n_3$, $n_4$ are non-negative integers,
and
\begin{eqnarray}
&&\xi:= \sqrt{2}r_0
\left(a-\cos\tilde\theta+{b\over 2}(\phi-\phi')^2\right)^{1/2}\,,
\\
\label{eq:xi}
&&R:=r-r_0\,,\quad\Theta:=\theta-{\pi\over2}\,,
\nonumber\\
\end{eqnarray}
with $a$, $b$ and $\phi'$ defined by
\begin{eqnarray}
a &:=&
1+{1\over 2r_0^2}{r_0^2\over r_0^2+{\cal L}^2}
{r_0^2\over(r_0-2M)^2}{\cal E}^2R^2
\,, \\
b &:=& {{\cal L}^2\over r_0^2}
\,, \\
\phi'&:=&-{{\cal L}\over r_0^2+{\cal L}^2}u_rR \,,
\end{eqnarray}
where ${\cal E}:=-g_{tt}{dt/d\tau}$, ${\cal L}:=g_{\phi\phi}{d\phi/d\tau}$,
and $u_r:=g_{rr}{dr/d\tau}$,
and $\tilde\theta$ is the relative angle
between $(\theta,\phi)$ and $(\pi/2,\phi')$.

There are two apparently different terms in
the covariant form of the direct force given by Eq.~(\ref{eq:dir_force});
the first term in the curly brackets exhibiting the quadratic divergence,
and the second term proportional to the curvature tensor that
appears to be finite in the coincidence limit.
In the local coordinate expansion, the second term will give terms
of the form $R/\xi$ or $\phi/\xi$.
As shown in Section~\ref{sec:comp}, the harmonic coefficients of $R/\xi$
vanish in the coincidence limit, while those of $\phi/\xi$
are finite but they give no contribution to the final result when
the infinite harmonic modes are summed up after
the coincidence limit is taken.
Hence we may focus on the first term. In passing, it is worthwhile
to note the following fact. Since the
direct force shows its dependence on the spin of the field
only through this second term (see Appendix~\ref{app:EMG}),
the harmonic coefficients of the direct force, which are to
be subtracted from the full force, will be independent of
the spin of the field.
\footnote{
This is also a result of the specific off-worldline extension 
chosen for the four-velocity. It is valid for the 
parallel-propagation extension, but does not hold, 
in general, for other extensions. }

Let us focus on the first term in the curly brackets of
Eq.~(\ref{eq:dir_force}). Since the orbit always remains
on the equatorial plane, the force is symmetric under the
transformation $\theta\to\pi-\theta$, which implies there is
no term proportional to odd powers of $\Theta$.
Hence we only need to consider the case of $n_2$ being an even
number in the general form given by Eq.~(\ref{eq:ingredient}).
Then the factor $\Theta^{n_2}$
may be eliminated by expressing $\Theta^2$ in terms of
$\xi$, $R$ and $\phi$, and we are left with terms of the form,
\begin{eqnarray}
{R^{n_1}\phi^{n_3} \over \xi^{2n_4+1}} \,.
\label{eq:dirform}
\end{eqnarray}
Explicitly, we find
\begin{eqnarray}
F_t^{\rm dir} &=& q\Biggl(
{\cal E} u_r{R\over\xi^3}+{\cal E}{\cal L}{\phi\over\xi^3}
\nonumber \\ && \qquad
-{1\over 2}{(r_0-2M){\cal E} u_r \over r_0^2}{1\over\xi}
+{2(r_0-2M){\cal E} {\cal L}^2 u_r \over r_0^2}{\phi^2\over\xi^3}
\nonumber \\ && \qquad
-{3\over 2}{(r_0-2M){\cal E} {\cal L}^4 u_r \over r_0^2}{\phi^4\over\xi^5}
\Biggr)
\,,
\label{eq:dirforcet}
\\
F_r^{\rm dir} &=& q\Biggl(
{{\cal L}^2\over r_0(r_0-2M)}{R\over\xi^3}
-{r_0^2{\cal E}^2\over (r_0-2M)^2}{R\over\xi^3}
-{\cal L}u_r{\phi\over\xi^3}
\nonumber \\ && \qquad
-{1\over 2}{2r_0^2+{\cal L}^2 \over r_0^3}{1\over\xi}
+{1\over 2}{{\cal E}^2 \over r_0-2M}{1\over\xi}
\nonumber \\ && \qquad
+{1\over 2}{(3r_0^2+4{\cal L}^2){\cal L}^2\over r_0^3}{\phi^2\over\xi^3}
-{2{\cal E}^2 {\cal L}^2 \over r_0-2M}{\phi^2\over\xi^3}
\nonumber \\ && \qquad
-{3\over 2}{(r_0^2+{\cal L}^2){\cal L}^4 \over r_0^3}{\phi^4\over\xi^5}
+{3\over 2}{{\cal E}^2 {\cal L}^4\over r_0-2M}{\phi^4\over\xi^5}
\,,
\label{eq:dirforcer}
\\
F_\theta^{\rm dir} &=& 0
\,,
\label{eq:dirforceth}
\\
F_\phi^{\rm dir} &=& q\Biggl(
-{\cal L} u_r {R\over\xi^3}-(r_0^2+{\cal L}^2){\phi\over\xi^3}
\nonumber \\ && \qquad
+{1\over 2}{(r_0-2M){\cal L} u_r \over r_0^2}{1\over\xi}
-{1\over 2}{(r_0-2M)(r_0^2+4{\cal L}^2){\cal L} u_r \over r_0^2}{\phi^2\over\xi^3}
\nonumber \\ && \qquad
+{3\over 2}{(r_0-2M)(r_0^2+{\cal L}^2){\cal L}^3 u_r \over r_0^2}
{\phi^4\over\xi^5}
\Biggr)
\,,
\label{eq:dirforceph}
\end{eqnarray}
where $F_{\alpha}^{\rm dir}=F_{\alpha}[\phi^{\rm dir}]$.
The absence of $F_\theta^{\rm dir}$ is because of the
symmetry of the background; the orbit remains on the equatorial
plane even under the action of the self-force.
In the above, we have discarded the terms
of the form $R/\xi$ or $\phi/\xi$. As mentioned before, and
as shown in Section~\ref{sec:comp},
such terms give no contribution to the final force.

What we have to do now is to perform the harmonic decomposition of the
components of the direct force given above.
To do so, we note the following important fact.
Apart from the trivial multiplicative factor of $R^{n_1}$ which
is independent of the spherical coordinates,
the terms to be expanded in the spherical harmonics are of
the form $\phi^{n_3}/\xi^{2n_4+1}$, or $(\phi-\phi')^{n_3}/\xi^{2n_4+1}$.
To the order of accuracy we need (in fact only the leading order
accuracy is sufficient; see Appendix~\ref{app:property}),
the factor $(\phi-\phi')^{n_3}$ may be eliminated by replacing
it to an equivalent $\phi$-derivative operator
of degree $n_3$ acting on $\xi^{2n_3-2n_4-1}$,
which is further converted to a polynomial in $m$ after
the harmonic expansion of $\xi^{2n_3-2n_4-1}$.
Thus the only basic formula we need is the harmonic expansion
of $\xi^{2p-1}$ where $p$ is an integer.
Detailed derivation of it is given in Appendix~\ref{app:formula}.
Note that, apart from the term $b(\phi-\phi')^2/2$ in $\xi$
with respect to which we expand $\xi$ in a convergent infinite series,
$\xi^{2p-1}$ is defined over the whole sphere to allow the straightforward
harmonic decomposition.
The result to the leading order in the coincidence limit $a \to1+0$ is
\begin{eqnarray}
\left({\xi\over\sqrt{2}\,r_0}\right)^{2p-1}
&&=\left(a-\cos\tilde\theta+{b\over 2}(\phi-\phi')^2\right)^{p-1/2}
= 2\pi\sum_{\ell=0}^{\infty}\sum_{m=-\ell}^{\ell}D^{p-1/2}_{\ell m}(a)
Y_{\ell m}(\theta,\phi)Y^*_{\ell m}(\theta',\phi')\,,
\label{eq:xipdec}
\\
D^{p-1/2}_{\ell m}(a) && \, \to
 \cases{ \displaystyle
{1\over \sqrt{1+b}}{1\over -p-1/2}(a-1)^{p+1/2} \,, &for $p+{1\over2}<0$,
\cr \displaystyle
{(-1)^\ell 2^{p+1/2} \over \sqrt{1+b}} \sum_{n=0}^\infty
{\Gamma(p+1/2) \Gamma(p+n+1/2)\over
\Gamma(p+n-\ell+1/2)\Gamma(p+n+\ell+3/2)}{1\over n!}
\left({-m^2 b\over 1+b}\right)^n  \,, & for $p+{1\over2}>0$\,. \cr}
\end{eqnarray}
We note that, although what we need here is only the case of an integer $p$,
the above formula is valid for any $p$ (except for the case $p=-1/2$).

After the decomposition, we can take the radial coincidence limit $r \to r_0$
(followed by the angular coincidence limit if desired).
The basic properties of the resulting mode coefficients
in the coincidence limit are discussed in Section~\ref{sec:comp}.
Here we briefly explain the reason why the terms proportional to
$R/\xi$ and $\phi/\xi$ give no contribution to the final
result.
The term $R/\xi$ corresponds to $R$ times the case of $p=0$, for which
$D_{\ell m}^{p-1/2}$ is finite in the limit $a\to1$
(i.e., $R\to0$). Hence all the coefficients vanish in the radial
coincidence limit.
As for $\phi/\xi$, it can be replaced by $(\phi-\phi')/\xi$
which is equivalent to $\partial_{\phi}\xi$ in the coincidence limit.
This corresponds to the case of $p=1$ multiplied by $m$.
Hence all the harmonic coefficients become odd functions of $m$,
and their sum over $m$ for each $\ell$ vanishes
in the angular coincidence limit.
As a result,
the non-vanishing contribution comes only from
the terms $R/\xi^3$, $\phi/\xi^3$,
$1/\xi$, $\phi^2/\xi^3$ and $\phi^4/\xi^5$.

\section{Regularization Counter Terms}\label{sec:comp}

In this section, we present the mode decomposition of
the direct force given by Eqs.~(\ref{eq:dirforcet})
$\sim$ (\ref{eq:dirforceph}), and compare the resulting
regularization counter terms with those obtained by
Barack and Ori\cite{BOnew} in their
{mode-sum regularization scheme} (MSRS)\cite{Barack}.

Barack and Ori define the regularization counter terms as
\begin{eqnarray}\label{ABC}
\lim_{x\to z_0}F_{\alpha l}^{\rm dir}=A_{\alpha}L+B_{\alpha}
+C_{\alpha}/L+O(L^{-2}) \,. \\
D_{\alpha}=
\sum_{l=0}^{\infty}\left[\lim_{x\to z_0}F_{\alpha l}^{\rm dir}
-A_{\alpha}L-B_{\alpha}-C_{\alpha}/L\right].
\end{eqnarray}
where $F_{\alpha l}^{\rm dir}$ is the multipole $l$-mode 
of $F_{\alpha}^{\rm dir}$, 
$L=\ell+1/2$, and $A_{\alpha}$, $B_{\alpha}$ and $C_{\alpha}$
are independent of $L$.
The $A_\alpha$ term is to subract the quadratic divergence,
the $B_\alpha$ term the linear divergence, and the $C_\alpha$
term the logarithmic divergence. The $D_\alpha$ term is
the remaining finite contribution of the direct force to
be subtracted. As shown in Appendix~\ref{app:property},
we find $C_\alpha=D_\alpha=0$ in agreement with
Barack and Ori\cite{Barack}. We also find
the complete agreement of $A_\alpha$ and $B_\alpha$ terms
with their results for a general geodesic orbit\cite{BMONS,BOnew}
as given below.

The direct part of the force to be considered
has the form given by Eq.~(\ref{eq:dirform}), which may be
re-written as
\begin{eqnarray}
{R^{n_1}(\phi-\phi')^{n_3}\over \xi^{2n_4+1}}\,,
\label{eq:dirform2}
\end{eqnarray}
where $(n_1,n_3,n_4)$ are non-negative integers. Because the highest
order of divergence is quadratic, it is sufficient to consider
the cases $n_1+n_3-2n_4=-1$, $0$ and $1$.\footnote{Although
we may further restrict $n_4$ to the range $0\leq n_4\leq2$ from the
explict form of the direct force in Eqs.~(\ref{eq:dirforcet}) $\sim$
(\ref{eq:dirforceph}),
we choose not to do so because it turns out to be unnecessary
in the following discussion.}

We first note that
\begin{eqnarray}
{\phi-\phi'\over \xi^{2n_4+1}} &=&
-{1\over 2n_4-1}{1\over r_0^2+{\cal L}^2}
{\partial\over\partial\phi}\left({1\over\xi^{2n_4-1}}\right)+O(y^{-2n_4+2})\,.
\label{eq:Pxig}
\end{eqnarray}
By using this equation recursively,
we obtain
\begin{eqnarray*}
{(\phi-\phi')^{n_3}\over\xi^{2n_4+1}}
\propto
{\partial^{n_3} \over \partial \phi^{n_3}} \xi^{2n_3-2n_4-1}
+O(y^{n_3-2n_4+1})\,.
\end{eqnarray*}
In the context of the harmonic decomposition, we may replace
the derivative $\partial/\partial\phi$ by $i\,m$.
Hence instead of Eq.~(\ref{eq:dirform2}), we may
consider the terms of the form,
\begin{eqnarray}
m^{n_3}R^{n_1}\xi^{2n_3-2n_4-1}+O(y^{n_1+n_3-2n_4+1})\,.
\label{eq:dirform3}
\end{eqnarray}

In the above equation, we have indicated by $O(y^{n_1+n_3-2n_4+1})$
the presence of correction terms of $O(y^2)$ relative to
the original form (\ref{eq:dirform2}).
In terms of the regularization parameters $A_\alpha$,
$B_\alpha$, $C_\alpha$ and $D_\alpha$, this implies
the terms with $n_1+n_3-2n_4=-1$
would contribute to $A_\alpha$ and $C_\alpha$, and
the terms with $n_1+n_3-2n_4=0$ to $B_\alpha$ and $D_\alpha$,
while the terms with $n_1+n_3-2n_4=1$ to $D_\alpha$.
However, as shown in Appendix~\ref{app:property}, by a general
argument, we can show that both $C_\alpha$ and $D_\alpha$ vanish.
Therefore we do not have to worry about the $O(y^2)$ corrections
in Eq.~(\ref{eq:dirform3}) but may focus on the leading behavior
of it in the coincidence limit.

Keeping this fact in mind, we now analyze which cases of the
form (\ref{eq:dirform3}) contribute to the regularization
parameters $A_\alpha$ and $B_\alpha$.
For this purpose, we set $n_1+n_3-2n_4=q$ where $q=-1$, $0$ or $1$.
Then comparing Eq.~(\ref{eq:dirform3}) with Eq.~(\ref{eq:xipdec}),
we find it is convenient to separately consider the two cases:
\begin{list}{}{}
\item[(1)] The case $2p=2n_3-2n_4=n_3-n_1+q\leq-2$.\\
In this case, the harmonic coefficients of Eq.~(\ref{eq:dirform3})
behave as
\begin{eqnarray*}
\sim R^{n_1+2n_3-2n_4+1}=R^{n_3+q+1}\,.
\end{eqnarray*}
Since $n_3\geq0$, the harmonic coefficients are non-vanishing in the
limit $R\to0$ only if $n_3=0$ and $q=-1$. This means
$n_1=2n_4-1$ ($\geq0$). Therefore only the terms of the form
$R^{2n_4-1}/\xi^{2n_4+1}$ ($n_4\geq 1$) give finite coefficients,
and they contribute to $A_\alpha$.

\item[(2)] The case $2p=2n_3-2n_4=n_3-n_1+q\geq0$.\\
In this case, since the harmonic coefficients of
$\xi^{2n_3-2n_4-1}$ are finite in the limit $R\to0$,
we must have $n_1=0$, hence $n_3=2n_4+q$. Therefore, since
$D_{\ell m}^{p-1/2}$ is an even function of $m$, the
harmonic coefficents will be odd functions of $m$ if $q$ is odd,
i.e., if $q=-1$ or $1$. When the sum over $m$ is taken,
the result vanishes in the angular coincidence limit if $q$ is odd
because of the symmetry property of
$|Y_{\ell m}(\theta,\phi)|^2$ under $m\to-m$.
Thus only the case of $q=0$ or $n_3=2n_4$ remains.
The corresponding terms
are of the form $(\phi-\phi')^{2n_4}/\xi^{2n_4+1}$, and
they contribute to $B_\alpha$.
\end{list}


{}From the above results, and noting that $\phi'\propto R$,
we obtain the equality,
\begin{eqnarray}
{(\phi-\phi')^{2n}\over\xi^{2n+1}}
={(\phi^2-2\phi\phi'+\phi'{}^2)^n\over\xi^{2n+1}}
={\phi^2\over\xi^{2n+1}}\,,
\end{eqnarray}
which holds in the sense of its contributions to
the regularization parameters. Thus, to summarize,
the non-vanishing contributions are from the terms
either of the form
$R^{2n+1}/\xi^{2n+3}$ or of the form $\phi^{2n}/\xi^{2n+1}$,
where $n$ is a non-negative integer, and the former
contributes to $A_\alpha$ while the latter to $B_\alpha$.

\subsection{The $A$-term}

The $A$-term describes the quadratic divergent terms of the direct force.
Thus we consider the most divergent terms in Eqs.~(\ref{eq:dirforcet})
$\sim$ (\ref{eq:dirforceph}),
\begin{eqnarray}
{R \over \xi^{3}} \qquad {\rm and} \qquad
{\phi \over \xi^{3}} \,.
\end{eqnarray}
As discussed above, the term $\phi/\xi^3$ may be replaced as
$(\phi-\phi')/\xi+\phi'/\xi=\phi'/\xi$. Hence we may focus on the
form $R/\xi^3$.
The essential fact is that this is odd in $R$.
This leads to the harmonic coefficients proportional to ${\rm sign}(R)$.
Using the fomula (\ref{eq:leading-mode0}), we obtain
\begin{eqnarray}
A_{t} &=& {\rm sign}(R) {q^2 \over r^2} {r_0-2M \over r_0}
{u_r \over 1-{\cal L}^2/r_0^2} \,,
\\
A_{r} &=& -{\rm sign}(R) {q^2 \over r^2} {r_0 \over r_0-2M}
{{\cal E} \over 1-{\cal L}^2/r_0^2} \,,
\\
A_{\phi} &=& 0 \,.
\end{eqnarray}
These $A$-terms vanish when averaged over both limits $R\to\pm0$.
There could be correction terms of $O(y^0)$ which could
contribute to the $C$ and $D$-terms.
However, as shown in Appendix~\ref{app:property},
they are known to be absent.

\subsection{The $B$-term}

The linearly divergent terms are described by the $B$-term, which
are of the form,
\begin{eqnarray}
{\phi^{2n} \over \xi^{2n+1}} \,,
\end{eqnarray}
in Eqs.~(\ref{eq:dirforcet}) $\sim$ (\ref{eq:dirforceph}).
The Legendre coefficients are given by the formula (\ref{eq:leading-mode1}).
We find
\begin{eqnarray}
B_t &=& -{(r_0-2M){\cal E}u_r \over 2r_0}
\sum_{n=0}^{\infty}{(2n)!\over 2^{2n} (n!)^2}(-1)^n
{(2n+1)^2\Gamma(n+1/2)\over \sqrt{\pi}\Gamma(n+1)}{{\cal L}^{2n}\over r_0^{2n+1}} \,,
\\
B_r &=& {(r_0-2M)u_r^2 \over 2r_0}
\sum_{n=0}^{\infty}{(2n)!\over 2^{2n} (n!)^2}
(-1)^n{(2n+1)^2\Gamma(n+1/2)\over \sqrt{\pi}\Gamma(n+1)}
{{\cal L}^{2n}\over r_0^{2n+1}}
\nonumber \\
&&- {1\over 2r_0}
\sum_{n=0}^{\infty}{(2n)!\over 2^{2n} (n!)^2}
(-1)^n{(-(2n-1)) \Gamma(n+1/2) \over \sqrt{\pi}\Gamma(n+1)}
{{\cal L}^{2n}\over r_0^{2n+1}} \,,
\\
B_{\phi} &=& {(r_0-2M){\cal L}u_r\over 2r_0^2}
\Biggl(\sum_{n=0}^{\infty}{(2n)!\over 2^{2n} (n!)^2}
(-1)^n{(2n+1)^2\Gamma(n+1/2)\over \sqrt{\pi}\Gamma(n+1)}
{{\cal L}^{2n}\over r_0^{2n+1}} \nonumber \\
&& \qquad \quad - {r_0^2\over {\cal L}^2}
\sum_{n=0}^{\infty}{(2(n+1))! \over 2^{2(n+1)}((n+1)!)^2}
(-1)^n{(2n+1)^2\Gamma(n+1/2)\over \sqrt{\pi}\Gamma(n+1)}
{{\cal L}^{2(n+1)}\over r_0^{2(n+1)+1}} \Biggr)\,.
\end{eqnarray}
The above may be expressed in terms of the hypergeometric functions
as
\begin{eqnarray}
B_t &=& -{(r_0-2M){\cal E}u_r \over 2r_0^3}
F\left({3\over2},{3\over2};1;-{{\cal L}^2\over r_0^2}\right) \,,
\\
B_r &=& {(r_0-2M)u_r^2 \over 2r_0^3}
F\left({3\over2},{3\over2};1;-{{\cal L}^2\over r_0^2}\right)
-{1\over 2r_0^2}
\left(
F\left({1\over2},{1\over2};1;-{{\cal L}^2\over r_0^2}\right)
+{{\cal L}^2 \over 2r_0^2}
F\left({3\over2},{3\over2};2;-{{\cal L}^2\over r_0^2}\right)
\right) \,,
\\
B_{\phi} &=& {(r_0-2M){\cal L}u_r\over 16r_0^3}
\left(
8F\left({3\over2},{3\over2};1;-{{\cal L}^2\over r_0^2}\right)
-4F\left({3\over2},{3\over2};2;-{{\cal L}^2\over r_0^2}\right)
+{9{\cal L}^2 \over r_0^2}
F\left({5\over2},{5\over2};3;-{{\cal L}^2\over r_0^2}\right)
\right) \,.
\end{eqnarray}

The above results for the $A$ and $B$-terms perfectly agree with
the results obtained by Barack and Ori in a quite different
fashion\cite{BMONS,BOnew}.
\footnote{
Here we give the values of $B_{\alpha}$ expressed 
in terms of generalized hypergeometric functions, 
while in \cite{BMONS} these are given in terms of 
the two complete elliptic integrals $K$ and $E$. 
The two expressions are related by changing of variables and 
using the formulas in \cite{HTF}.}

\section{Conclusion and Discussion}\label{sec:concl}

Our final goal is to establish a method of calculation of
the local gravitational reaction force to a point particle
orbiting a Kerr black hole.
We have pointed out in Section \ref{sec:intro}
that there are two problems;
the `subtraction problem' and the `gauge problem'.

In this paper,
we have only discussed a possible approach to the subtraction problem.
We have introduced a regularization method which utilizes
the spherical-harmonic decomposition, and have derived the direct part
of the self-force, which turns out to be
independent of the spin $s$ of the field under consideration.
The harmonic decomposition of this direct part has been carried out,
and the regularization counter terms for the self-force
have been derived for a general geodesic orbit.
We have found our result agrees completely with the result obtained by
Barack and Ori\cite{BOnew} in their mode-sum regularization
scheme (MSRS) \cite{Barack}.

To compare with the MSRS, we have derived the regularization counter
terms which are obtained by summating the harmonic coefficients over $m$.
However, when we extend our method to the Kerr background,
we may have to carry out the regularization before taking the
$m$-summation. In this sense, the formulas derived in
Appendix~\ref{app:formula}, where no summation over $m$ is assumed,
may be still useful in the Kerr case.

It is worthwhile to point out that the gauge problem in the
gravitational case seems far more serious than the subtraction problem.
What we know at the moment is that the gravitational self-force
is described by the tail part of the metric perturbation induced
by a particle\cite{reaction,QuinnWald}.
However this is justified only in the harmonic gauge,
while the full metric perturbation can be obtained
only in the Regge-Wheeler gauge or in the radiation gauge
where the identification of the tail part is highly non-trivial.
A prescription to identify the tail part of the metric perturbation
has been proposed in \cite{NakSas}, but it needs to be verified.
The gauge problem for the non-radiative monopole and
dipole components of the metric perturbation which are not obtainable
in the Teukolsky formalism seems to stand as
additional serious obstacle.
Possible resolutions for the gauge problem are under investigation.
\footnote{In this respect, recently some progress has been made in 
\cite{BarOri2}.}

\section*{Acknowledgements}
We would like to thank L.~Barack, A.~Ori and H.~Tagoshi
for useful discussions.
YM and HN are grateful to K.~Thorne for his hospitality
and discussions during their stay at CalTech,
and thank all the participants at the 3rd Capra Ranch Meeting,
Caltech in US for discussions.
We also thank all the participants at the 4rd Capra Ranch Meeting,
Golm in Germany, for invaluable discussions.
YM is grateful B.~Schutz for his hospitality and discussions
during his stay at the Einstein Institute at Potsdam.

This work was supported in part by a Monbusho Grant-in-Aid
for Creative Research (No.~09NP0801), and by
a Monbusho Grant-in-Aid for Scientific Research (No.~12640269).
YM was supported by NSF Grant PHY-0099568, PHY-0096522 and
NASA Grant NAG5-10707.
HN is supported by Research Fellowships of the
Japan Society for the Promotion of Science
for Young Scientists, No.~2397.


\begin{appendix}

\section{Bi-tensors and local coordinate expansion}\label{app:bi-tensor}

Bi-tensors are tensors which depend on two distinct field points,
say, $x^\alpha$ and $\bar x^{\bar \alpha}$.
For our purpose, we consider
half the squared geodesic interval bi-scalar $\sigma(x,\bar x)$,
and the geodesic paralell displacement bi-vector
${\bar g}_{\alpha {\bar \alpha}}(x,\bar x)$,
which satisfy
\begin{eqnarray}
&& \sigma(x,{\bar x})
~=~ {1\over 2}g^{\alpha\beta}
\sigma_{;\alpha}(x,{\bar x})\sigma_{;\beta}(x,{\bar x})
~=~ {1\over 2}g^{{\bar \alpha}{\bar \beta}}
\sigma_{;{\bar \alpha}}(x,{\bar x})\sigma_{;{\bar \beta}}(x,{\bar x}) \,,
\\
&& \lim_{x\rightarrow{\bar x}}\sigma_{;\alpha}(x,{\bar x})
~=~ \lim_{x\rightarrow{\bar x}}\sigma_{;{\bar \alpha}}(x,{\bar x})
~=~ 0 \,,
\\
&& {\bar g}_{\alpha{\bar \alpha};\beta}(x,{\bar x})
g^{\beta\gamma}(x)\sigma_{;\gamma}(x,{\bar x}) ~=~ 0 \,,\quad
{\bar g}_{\alpha{\bar \alpha};{\bar \beta}}(x,{\bar x})
g^{{\bar \beta}{\bar \gamma}}({\bar x})
\sigma_{;{\bar \gamma}}(x,{\bar x}) ~=~ 0 \,,
\\
&& \lim_{x\rightarrow{\bar x}}{\bar g}_\alpha{}^{\bar \alpha} ~=~
\delta_\alpha{}^{\bar \alpha} \,.
\end{eqnarray}
We also need
the generalized van Vleck-Morette determinant bi-scalar,
\begin{eqnarray}
\Delta(x,{\bar x}) &=&
\det\bigl(-{\bar g}^{\alpha{\bar\alpha}}(x,{\bar x})
\sigma_{;{\bar\alpha}\beta}(x,{\bar x})\bigr) \,.
\end{eqnarray}
We consider the local expansion of these bi-tensors
around the coincidence limit $x\rightarrow {\bar x}$.

In calculating the local expansion of the field and its derivatives
in a covariant way,
the following formula are useful\cite{reaction}.
\begin{eqnarray}
\sigma_{;\alpha\beta}(x,z) &=& g_{\alpha\beta}(z)- {1\over 3}
R_{\alpha}{}^{\gamma}{}_{\beta}{}^{\delta}(z)
\sigma_{;\gamma}(x,z) \sigma_{;\delta}(x,z) +O(\epsilon^3) \,,
\label{128db} \\
\sigma_{;\mu\beta}(x,z) &=& -\bar g_{\mu}{}^{\alpha}(x,z)
\left(g_{\alpha\beta}(z)+{1\over 6}
R_{\alpha\gamma\beta\delta}(z)
\sigma^{;\gamma}(x,z) \sigma^{;\delta}(x,z)\right)
\cr &&
+O(\epsilon^3) \,,
\label{173db} \\
\bar g^{\mu\alpha}{}_{;\beta}(x,z) & = & -{1\over 2}
\bar g^{\mu\gamma}(x,z)R^{\alpha}{}_{\gamma\beta\delta}(z)
\sigma^{;\delta}(x,z) +O(\epsilon^2) \,, \cr
\bar g^{\mu\alpha}{}_{;\nu}(x,z) & = & -{1\over 2}
\bar g^{\mu\beta}(x,z) \bar g_{\nu}{}^{\gamma}(x,z)
R^{\alpha}{}_{\beta\gamma\delta}(z)
\sigma^{;\delta}(x,z) +O(\epsilon^2) \,.
\label{140db}
\end{eqnarray}
As for the calculation of $\Delta(x,{\bar x})$,
we use the result of the covariant expansion given in Ref.\cite{DeWittBrehme}.
We have
\begin{eqnarray}
\sigma_{;{\bar\alpha}\beta}(x,{\bar x}) &=&
-{\bar g}_\beta{}^{\bar \beta}(x,{\bar x})
\left(g_{\bar\alpha\bar\beta}({\bar x})
+{1\over 6}R_{{\bar\alpha}{\bar\gamma}{\bar\beta}{\bar\delta}}({\bar x})
\sigma^{;{\bar\gamma}}(x,{\bar x})
\sigma^{;{\bar\delta}}(x,{\bar x})
+O(|x-{\bar x}|^3)\right) .
\end{eqnarray}
Then, for a vacuum background, we have
\begin{eqnarray}
\Delta(x,{\bar x}) &=& 1 +O(|x-{\bar x}|^3) \,.
\end{eqnarray}

The local expansion of these bi-tensors by the background coordinates
is derived from the formulas (\ref{eq:sigma-exp}) and (\ref{eq:g-exp}),
and the expansion coefficiets expressed in terms of
the ordinary derivatives of the background metric
are given in Eqs.~(\ref{eq:A3}), (\ref{eq:A4}), (\ref{eq:B3}) and
(\ref{eq:B4}).
In the Schwarzschild background,
the non-vanishing components of these coefficients are
\begin{eqnarray}
&& A_{ttr}=A_{trt}=A_{rtt}=-{M\over r^2} \,, \nonumber\\
&& A_{rrr}=-{3M\over(r-2M)^2} \,, \nonumber\\
&& A_{r\theta\theta}=A_{\theta r\theta}=A_{\theta\theta r}=r \,, \nonumber\\
&& A_{r\phi\phi}=A_{\phi r\phi}=A_{\phi\phi r}=r\sin^2\theta \,, \nonumber\\
&& A_{\theta\phi\phi}=A_{\phi\phi\theta}=A_{\phi\theta\phi}
=r^2\sin\theta\cos\theta \,, \nonumber\\
&& A_{tttt}=-{M^2(r-2M)\over r^5} \,, \nonumber\\
&& A_{ttrr}=A_{trtr}=A_{rttr}=A_{trrt}=A_{rtrt}=A_{rrtt}
={M(4r-5M)\over 3r^3(r-2M)} \,, \nonumber\\
&& A_{tt\theta\theta}=A_{t\theta t\theta}=A_{\theta tt\theta}
=A_{t\theta\theta t}=A_{\theta t\theta t}=A_{\theta\theta tt}
={M(r-2M)\over 3r^2} \,, \nonumber\\
&& A_{tt\phi\phi}=A_{t\phi t\phi}=A_{\phi tt\phi}=A_{t\phi\phi t}
=A_{\phi t\phi t}=A_{\phi\phi tt}
={M(r-2M)\over 3r^2}\sin^2\theta \,, \nonumber\\
&& A_{rrrr}={M(8r-M)\over r(r-2M)^3} \,, \nonumber\\
&& A_{rr\theta\theta}=A_{r\theta r\theta}=A_{\theta rr\theta}
=A_{r\theta\theta r}=A_{\theta r\theta r}=A_{\theta\theta rr}
=-{M\over 3(r-2M)} \,, \nonumber\\
&& A_{rr\phi\phi}=A_{r\phi r\phi}=A_{\phi rr\phi}=A_{r\phi\phi r}
=A_{\phi r\phi r}=A_{\phi\phi rr}
=-{M\over 3(r-2M)}\sin^2\theta \,, \nonumber\\
&& A_{\theta\theta\theta\theta}=-r(r-2M) \,, \nonumber\\
&& A_{\theta\theta\phi\phi}=A_{\theta\phi\theta\phi}=A_{\phi\theta\theta\phi}
=A_{\theta\phi\phi\theta}=A_{\phi\theta\phi\theta}=A_{\phi\phi\theta\theta}
=-{r(3r-2M)\over 3}\sin^2\theta \,, \nonumber\\
&& A_{\phi\phi\phi\phi}
=-r(r-2M)\sin^4\theta-r^2\sin^2\theta\cos^2\theta \,, \nonumber\\
&& A_{r\theta\phi\phi}=A_{r\phi\theta\phi}=A_{r\phi\phi\theta}
=A_{\theta r\phi\phi}=A_{\phi r\theta\phi}=A_{\phi r\phi\theta}
=A_{\theta\phi r\phi} \nonumber \\ && \quad
=A_{\phi\theta r\phi}=A_{\phi\phi r\theta}=A_{\theta\phi\phi r}
=A_{\phi\theta\phi r}=A_{\phi\phi\theta r}
= r\sin\theta\cos\theta \,,
\nonumber\\
&& B_{tt|r}=-B_{tr|t}=B_{rt|t}=-{M\over r^2} \,, \nonumber\\
&& B_{rr|r}=-{M\over (r-2M)^2} \,, \nonumber\\
&& B_{\theta\theta|r}=-B_{\theta r|\theta}=B_{r\theta|\theta}=r \,,
\nonumber\\
&& B_{\phi\phi|r}=-B_{\phi r|\phi}=B_{r\phi|\phi}=r\sin^2\theta \,,
\nonumber\\
&& B_{\theta\phi|\phi}=-B_{\phi\theta|\phi}=B_{\phi\phi|\theta}
=r^2\sin\theta\cos\theta \,, \nonumber\\
&& B_{tt|tt}=-{M^2(r-2M)\over r^5} \,, \nonumber\\
&& B_{tt|rr}={M(2r-3M)\over r^3(r-2M)} \,, \nonumber\\
&& B_{tr|tr}=B_{tr|rt}=-{M(r-3M)\over r^3(r-2M)} \,, \nonumber\\
&& B_{rt|tr}=B_{rt|rt}={M(r-M)\over r^3(r-2M)} \,, \nonumber\\
&& B_{rr|tt}={M^2\over r^3(r-2M)} \,, \nonumber\\
&& B_{t\theta|t\theta}=B_{t\theta|\theta t}
=B_{\theta t|t\theta}=B_{\theta t|\theta t}
={M(r-2M)\over 2 r^2} \,, \nonumber\\
&& B_{t\phi|t\phi}=B_{t\phi|\phi t}=B_{\phi t|t\phi}=B_{\phi t|\phi t}
={M(r-2M)\over 2 r^2}\sin^2\theta \,, \nonumber\\
&& B_{rr|rr}={M(2r-M)\over r(r-2M)^3} \,, \nonumber\\
&& B_{rr|\theta\theta}=-1 \,, \nonumber\\
&& B_{r\theta|r\theta}=B_{r\theta|\theta r}=-{M\over 2(r-2M)} \,, \nonumber\\
&& B_{\theta r|r\theta}=B_{\theta r|\theta r}=-{2r-3M\over 2(r-2M)}
\,, \nonumber\\
&& B_{rr|\phi\phi}=-\sin^2\theta \,, \nonumber\\
&& B_{r\phi|r\phi}=B_{r\phi|\phi r}=-{M\over 2(r-2M)}\sin^2\theta \,,
\nonumber\\
&& B_{\phi r|r\phi}=B_{\phi r|\phi r}=-{2r-3M\over 2(r-2M)}\sin^2\theta \,,
\nonumber\\
&& B_{\theta\theta|\theta\theta}=-r(r-2M) \,, \nonumber\\
&& B_{\theta\theta|\phi\phi}=-r^2\cos^2\theta \,, \nonumber\\
&& B_{\theta\phi|\theta\phi}=B_{\theta\phi|\phi\theta}
=-r(r-M)\sin^2\theta \,, \nonumber\\
&& B_{\phi\theta|\theta\phi}=B_{\phi\theta|\phi\theta}
=-r^2\cos^2\theta+Mr\sin^2\theta \,,\nonumber \\
&& B_{\phi\phi|\theta\theta}=-r^2\sin^2\theta \,,\nonumber \\
&& B_{\phi\phi|\phi\phi}
=-r(r-2M)\sin^4\theta-r^2\sin^2\theta\cos^2\theta \,, \nonumber\\
&& B_{r\theta|\phi\phi}=B_{\theta r|\phi\phi}
=-B_{\theta\phi|r\phi}=-B_{\theta\phi|\phi r}
=B_{\phi\theta|r\phi}=B_{\phi\theta|\phi r}
=-B_{r\phi|\theta\phi} \nonumber \\ && \quad
=-B_{r\phi|\phi\theta}
=B_{\phi r|\theta\phi}=B_{\phi r|\phi\theta}
=-B_{\phi\phi|r\theta}=-B_{\phi\phi|\theta r}
=-r\sin\theta\cos\theta \,. \nonumber
\end{eqnarray}

\section{The direct part of electromagnetic and gravitational self-force}
\label{app:EMG}

In this appendix, we summarize the direct part of the vector and
tensor field.
The direct part of the field is obtained
by integrating the direct part of the Green function ${}_sG_{\{A\}}^{\rm dir}$
as same as the scalar case.
\begin{eqnarray}
{}_sG_{\{A\}}^{\rm dir}(x,x') &=& -{1\over 4\pi}\, \theta[\Sigma(x),x']
{}_su_{\{A\}}(x,x')\delta\bigl(\sigma(x,x')\bigr)
\,, \label{eq:EMGdirect-g} \\
{}_su_{\{A\}}(x,x') &=& \cases{
\displaystyle \sqrt{\Delta(x,x')} & ($s=0$), \cr
\displaystyle \sqrt{\Delta(x,x')}\bar g_{\mu\mu'}(x,x')  & ($s=1$), \cr
\displaystyle \sqrt{\Delta(x,x')}\bar g_{\mu\mu'}(x,x')\bar g_{\nu\nu'}(x,x')
& ($s=2$), }
\end{eqnarray}
where the suffix $\{A\}$ stands for the spacetime indices appropriate to
the spin $s$ of the field.

From the above Green functions, we obtain
the direct part of the field which is expanded
around the particle location as
\begin{eqnarray}
{}_s\phi_{\{A\}}^{\rm dir}(x) &=& \cases{
\displaystyle q\left[{1 \over \sigma_{;\alpha}(x,z(\tau_{\rm ret}))
v^\alpha(\tau_{\rm ret})}\right] +O(y^2) \,, & scalar \,, \cr
\displaystyle e\left[{
\bar g_{\mu\alpha}(x,z(\tau_{\rm ret}))v^\alpha(\tau_{\rm ret})
\over \sigma_{;\beta}(x,z(\tau_{\rm ret}))v^\beta(\tau_{\rm ret})}
\right] +O(y^2) \,, & vector \,, \cr
\displaystyle 4Gm\left[{
\bar g_{\mu\alpha}(x,z(\tau_{\rm ret}))\bar g_{\nu\beta}(x,z(\tau_{\rm ret}))
v^\alpha(\tau_{\rm ret})v^\beta(\tau_{\rm ret})
\over \sigma_{;\gamma}(x,z(\tau_{\rm ret}))v^\gamma(\tau_{\rm ret})}
\right] +O(y^2) \,, & tensor \,,}
\label{eq:EMGdirect}
\end{eqnarray}
And then, we have
\begin{eqnarray}
F_\alpha[{}_s\phi_{\{A\}}^{\rm dir}](x) &=&
c {\bar g}_\alpha{}^{\bar \alpha}(x,z_{\rm eq}){1\over \epsilon^3\kappa}
\left\{\sigma_{;\bar\alpha}(x,z_{\rm eq})
+h\epsilon^2R_{\bar\alpha\bar\beta\bar\gamma\bar\delta}(z_{\rm eq})
v^{\bar\beta}_{\rm eq}\sigma^{;\bar\gamma}(x,z_{\rm eq})
v^{\bar\delta}_{\rm eq}\right\}+O(y)
\,, \label{eq:EMGdir_force} \\
\epsilon &=& \sqrt{2\sigma(x,z_{\rm eq})}
\,, \\
\kappa &=& \sqrt{-\sigma_{\bar\alpha\bar\beta}(x,z_{\rm eq})
v^{\bar\alpha}_{\rm eq}v^{\bar\beta}_{\rm eq}}
\nonumber \\
&=& 1+{1\over 6}R_{\bar\alpha\bar\beta\bar\gamma\bar\delta}(z_{\rm eq})
v^{\bar\alpha}_{\rm eq}\sigma^{;\bar\beta}(x,z_{\rm eq})
v^{\bar\gamma}_{\rm eq}\sigma_{;\bar\delta}(x,z_{\rm eq}) +O(y^3)
\,,
\end{eqnarray}
where $c$ and $h$ depenend on the spin of the field as
\begin{eqnarray}
(c\,,~h) &=&\displaystyle\cases{
(q^2\,,~1/3) \,, & scalar \,, \cr
(-e^2\,,~-2/3) \,, & vector \,, \cr
(Gm^2\,,~-11/3) \,, & tensor \,.}
\end{eqnarray}
Here an extension of the four velocity $v^{\alpha}(\tau_0)$
necessary to define the projection tensor $P_{\alpha}{}^{\beta}$,
as mentioned in the line following Eq.~(\ref{eq:force}),
is chosen such that
\begin{eqnarray}
V^{\alpha}(x)={\bar g}^\alpha{}_{\bar \alpha}(x,z_{\rm eq})
v^{\bar \alpha}(\tau_{\rm eq}(x)) \,.
\end{eqnarray}

It is noted that when we consider the mode decomposition
regularization for the self-force,
the direct part calculated in Eqs.
(\ref{eq:dirforcet})$\sim$(\ref{eq:dirforceph})
is independent of spin.

\section{Basic properties of the mode coefficients}\label{app:property}

In this appendix,
we examine the general properties of the mode coefficients
for the terms that appear in the local coordinate expansion
of the direct force given
in Eqs.~(\ref{eq:dirforcet}) $\sim$ (\ref{eq:dirforceph}).
We show that the $C$ and $D$-terms of the regularization couter
terms vanish in the coincidence limit to the particle position.

We first express $\xi$ in the form,
\begin{eqnarray}
\xi^2 &=&
\xi_0^2 + {\cal L}^2\left(\phi-\phi'\right)^2\,,
\end{eqnarray}
where $\xi_0$ is defined by
\begin{eqnarray}
\xi_0=\sqrt{2}r_0(a-\cos\tilde\theta)^{1/2} \,.
\end{eqnarray}
In terms of $\xi_0$, $R$ and $\phi$, all the terms that contribute to the
direct force $F_{\alpha}^{\rm dir}$ will have the form,
\begin{eqnarray}
F_{\alpha}^{\rm dir} \sim
{R^p(\phi-\phi')^{q}\xi_0^{2r}\over\xi_0^3}
\left({R^2\over\xi_0^2}\right)^m
\left({\phi^2\over\xi_0^2}\right)^n,
\label{generalform}
\end{eqnarray}
where we have replaced possible factors of the form $\Theta^{2k}$
in favor of polynomials in $\xi_0^2$, $R$ and $\phi$,
and $m$, $n$, $p$, $q$ and $r$ are non-negative integers
satisfying
\begin{eqnarray}
&&m\geq0\,,\quad n\geq0\,,
\quad
1\leq p+q+2r\leq3\,.
\end{eqnarray}
This is because the highest order of the divergence in the direct force
is quadratic, and we may focus only on terms of order
up to $O(y^0)$ in the local coordinate expansion.

Let us analyze the coincidence limit in detail.
By using a modified version of Eq.~(\ref{eq:Pxig}) with ${\cal L}^2=0$
but by taking account of $O(y^2)$ corrections,
one can further reduce the above to the form,
\begin{eqnarray}
(\partial_\phi)^{q+2n}\xi_0^{2q+2n+2r-2m-3}R^{2m+p}\,.
\label{genform}
\end{eqnarray}
Note that the $O(y^2)$ corrections only change the original $q$
to $q+2$, hence it is enough to consider the above form.

We can decompose (\ref{genform}) into a spherical harmonic series
by using the formula,
\begin{eqnarray}
C_\ell={2\ell+1\over2}\int_{-1}^1{P_\ell(\mu)\over\sqrt{a-\mu}} d\mu=
\sqrt{2}\left(a-\sqrt{a^2-1}\right)^{\ell+1/2}.
\label{Cell}
\end{eqnarray}
This $C_\ell$ is equal to the special case of
$C_\ell^p(a)$ with $p=-1/2$ given by Eq.~(\ref{eq:Cellp}).
Introducing a variable $z$ by $e^{z}=a+\sqrt{a^2-1}$ (hence $z>0$),
Eq.~(\ref{Cell}) is re-expressed as
\begin{eqnarray}
C_\ell(z)={2\ell+1\over2}\int_{-1}^1
{P_\ell(\mu)\over\sqrt{\cosh z-\mu}} d\mu=
e^{-Lz}\,,
\end{eqnarray}
where $L=\ell+1/2$. Note that $z\propto|R|$.

As discussed in Section~\ref{sec:comp}, odd powers of the operator
$\partial_\phi$
will give the harmonic coefficients which vanish after summing over $m$.
Hence we only need to consider the case of $q$ being an even non-negative
integer. Then it is straightforward to see that
the operator $\partial_\phi^2$
in the harmonic decomposition of (\ref{genform}) will give the factor
$L^2$ in the coincidence limit of the angular coordinates followed
by the sum over $m$.
In general,  $\partial_\phi^{2n}$ will give rise
to a polynomial of degree $n$ in $L^2$.
Explicitly, we have
\begin{eqnarray}
\sum_{m=-\ell}^{\ell} m^{2n} \left|Y_{\ell m}\left({\pi \over 2},0\right)
\right|^2
= {L \over 2\pi}\sum_{p=0}^n \lambda_p^{(n)} L^{2 p} \,.
\label{eq:Mn}
\end{eqnarray}
where the factor $\lambda_p^{(n)}$ is independent of $L$ and
$\lambda_n^{(n)}=\Gamma(n+1/2)/(\sqrt{\pi}\, \Gamma(n+1))$.
The derivation of this formula is given in Appendix C
of Ref.~\cite{NakanoMinoSasaki}.

Thus we may replace $(\partial_\phi)^{q+2n}$ in (\ref{genform}) by
$L^{2j}$ ($1\leq j\leq q/2+n$; note that $q$ is even).
With this replacement, let us consider the following 3
cases for the powers of $\xi_0$ separately:
\begin{list}{}{}
\item[(1)] $2N+1:=2m+3-(2q+2r+2n)>1$ ($\xi_0^{-(2N+1)}$; $N\geq1$).\\
In this case, to obtain
the harmonic decomposion of $\xi_0^{-(2N+1)}$,
one simply applies $[d/d(\cosh z)]^N=[d/\sinh zdz]^N$ to it:
\begin{eqnarray}
 \int_{-1}^1{P_\ell(\mu)\over(\cosh z-\mu)^{(2n+1)/2}} d\mu
\propto
\left[{d\over\sinh zdz}\right]^ne^{-Lz}\,.
\end{eqnarray}
Taking account of the general form (\ref{genform}),
this will give rise to the harmonic coefficients as
\begin{eqnarray}
\sim R^{2m+p}L^{2j}
{L^{k}\over z^{2N-k}}(1+\sum_{i\geq1}c_iz^{2i})e^{-Lz},
\end{eqnarray}
where $1\leq k\leq N$. Since $z\sim |R|$ and
\begin{eqnarray*}
2m+p&=&2N+2q+2r+2n-2+p
\\
&\geq& 2N+q+(q+2r+p)-2
\\
&\geq& 2N+q-1\geq 2N-1\,,
\end{eqnarray*}
the only term that remains in the limit $R\sim \pm z\to0$ is
the $k=1$ term, and this implies $2m+p$ is odd. Since we know
the leading divergence is quadratic, this will give
a harmonic coefficient proportional to ${\rm sign}(R)\,L$
which contributes to $A_\mu$.

\item[(2)] $2m+3-(2q+2r+2n)=1$ ($\xi_0^{-1}$).\\
In this case, the harmonic coefficents are non-vanishing only if
$2m+p=0$. Since the result is independent of $L$, it
contributes to $B_\mu$, i.e., the linearly divergent term.

\item[(3)] $2N+1:=2q+2r+2n-(2m+3)>1$ ($\xi_0^{2N+1}$; $N\geq1$).\\
In this case, since the harmonic coefficients of $\xi_0^{2N+1}$
will be finite in the limit $z\to0$, we must have $m=p=0$ which implies
$q+2r=2$.
Going back to the original form (\ref{genform}), one can then redefine $n$
by $n+q/2$ and the term of our interest takes the form,
\begin{eqnarray}
\partial_\phi^{2n}\xi_0^{2n-1}\,\quad (n\geq1).
\end{eqnarray}
Using the formula (\ref{eq:leading-mode0}),
we obtain the coefficients of the Legendre
decomposition of $\xi_0^{2n-1}$ as
\begin{eqnarray}
\xi_0^{2n-1}\Bigl|_L
={\kappa_{n}\over(L^2-1^2)(L^2-2^2)\cdot\cdot\cdot(L^2-n^2)}\,,
\label{eq:En}
\end{eqnarray}
where $\kappa_{n}=(-1)^n\left[ (2n-1)!! \right] r_0^{2n-1}$,
and we have introduced the notation,
\begin{eqnarray*}
{\cdots}~\Bigl|_L\,,
\end{eqnarray*}
to denote the $L=\ell+1/2$ mode coefficient of the
Legendre expansion in the coincidence limit.\footnote{When
the $m$-sum is non-trivial,
$\cdot\cdot\cdot|_L$ is defined to be the coefficient
after summating over $m$.}
Therefore, together with Eq.~(\ref{eq:Mn}), we have
\begin{eqnarray}
\partial_{\phi}^{2n} \xi_0^{2n-1}\Bigl|_{L}
&=&
{(-1)^n \kappa_{n}\over(L^2-1^2)(L^2-2^2)\cdot\cdot\cdot(L^2-n^2)}
\sum_{p=0}^n \lambda_p^{(n)} L^{2 p}
\,, \nonumber \\
&=& (-1)^n \kappa_{n}\,\lambda_{n}^{(n)}
+ \sum_{p=0}^{n-1}
{\nu_{p}\over(L^2-1^2)(L^2-2^2)\cdot\cdot\cdot(L^2-(n-p)^2)} \,,
\label{Lexp}
\end{eqnarray}
where $\nu_{p}$ is independent of $L$. We thus find
the first term on the right hand side, which is independent of $L$,
contributes to $B_\mu$, while the rest seem to give non-vanishing
contributions to $D_\mu$. However, Eq.~(\ref{eq:En}) tells us that
they are simply the Legendre coefficients of positive powers of
$\xi_0$, hence they vanish after the sum over $\ell$ is taken.
\end{list}
In all of the above three cases, nothing contributes to $C_\mu$.
Thus, to summarize, we find the $C$ and $D$-terms vanish and
the $A$-term is proprotional to ${\rm sign}(R)$.

\section{Mathematical formulas for mode decomposition}\label{app:formula}

In this appendix, we give formulas
neccesary for the harmonic decomposition of $\xi^{2n-1}$.
For this purpose, we introduce a dimensionless version of
$\xi$ by
\begin{eqnarray*}
\tilde\xi
=\left(a-\cos\tilde\theta+{b\over2}(\phi-\phi')^2\right)^{1/2}\,,
\end{eqnarray*}
Here, as defined in the text, $\tilde\theta$ is the angle between
$\{\theta,\,\phi\}$ and $\{\theta',\,\phi'\}$,
and $a>1$. In the following, we
do not restrict $n$ to be an integer. Our strategy is
to consider a series expansion of $\tilde\xi$
by treating $b$ as the expansion parameter.

First, we consider the harmonic expansion of
$\tilde\xi_0^{2p}$ where $\tilde\xi_0=(a-\cos\tilde\theta)^{1/2}$,
\begin{eqnarray}
\tilde\xi_0^{2p}=(a-\cos\tilde\theta)^p &=&
2\pi\sum_{\ell=0}^{\infty}\sum_{m=-\ell}^{\ell}C^p_\ell(a)
Y_{\ell m}(\theta,\phi)Y^*_{\ell m}(\theta',\phi')\,.
 \label{eq:mainterm0}
\end{eqnarray}
The expansion coefficients $C^p_\ell$ are obtained
by the Legendre integration as
\begin{eqnarray}
C^p_\ell(a) &=& \int^1_{-1}d\mu (a-\mu)^p P_\ell(\mu) \,,
\end{eqnarray}
where $P_\ell(\mu)$ is the Legendre function of the first kind.
As $a>1$, we can expand $(a-\mu)^p$ in powers of $\mu$.
Then we obtain
\begin{eqnarray}
C^p_\ell(a)
&=& {1\over 2^{p+1}\Gamma(-p)}a^{p-\ell}
\sum_{k=0}^{\infty}{\Gamma(k-p/2+\ell/2)
\Gamma(k-p/2+\ell/2+1/2)\over \Gamma(k+\ell+3/2)}
{1\over k!}\left({1\over a^2}\right)^k
\nonumber\\
&=& \sqrt{\pi}2^{-\ell}a^{p-\ell}
{\Gamma(-p+\ell)\over\Gamma(-p)\Gamma(\ell+3/2)}
F\left({-p+\ell \over 2},{-p+\ell+1 \over 2},\ell+{3\over 2};
{1\over a^2}\right)
\nonumber\\
&=& -{1\over p+1}a^{-\ell-1}\left({a^2-1\over 2a}\right)^{p+1}
F\left({p \over 2}+{\ell \over 2}+1,{p \over 2}
+{\ell \over 2}+{3 \over 2},p+2;1-{1 \over a^2}\right)
\nonumber \\ &&
+(-1)^\ell 2^{p+1}a^{p-\ell}
{\Gamma(p+1)^2\over \Gamma(p-\ell+1)\Gamma(p+\ell+2)}
F\left(-{p \over 2}+{\ell \over 2},
-{p \over 2}+{\ell \over 2}+{1 \over 2},-p;1-{1 \over a^2}\right) \,.
\label{eq:Cellp}
\end{eqnarray}
In the coincidence limit $a\to 1+0$,
we have the two qualitatively different leading behaviors
depending on the value of $p$ as
\begin{eqnarray}
C^p_\ell(a) \to \cases{ \displaystyle
{1\over -p-1}(a-1)^{p+1}, &for $p+1<0$,\cr
(-1)^\ell 2^{p+1} \displaystyle
{\Gamma(p+1)^2\over \Gamma(p-\ell+1)\Gamma(p+\ell+2)},
&for $p+1>0$\,. \cr}
\label{eq:leading-mode0}
\end{eqnarray}
It should be noted that the divergent behavior
persists in the mode coefficients for $p+1<0$.

Next, we consider the harmonic expansion of
$(\phi-\phi')^n(a-\cos\tilde\theta)^p$.
We take into account only the leading order behavior in the
coincidence limit.
Hence we have the basic formula which converts a power of
$(\phi-\phi')$ to the $\phi$-derivatives,
\begin{eqnarray}
(\phi-\phi')(a-\cos\tilde\theta)^p &=&
{1\over p+1}{\partial \over \partial\phi}(a-\cos\tilde\theta)^{p+1}
+ O(y^{2p+3}) \,. \label{eq:basic-rec}
\end{eqnarray}
Using this formula reccursively, we obtain
\begin{eqnarray}
(\phi-\phi')^n(a-\cos\tilde\theta)^p &=& \sum_{k=0}^{[n/2]} a^{(n)}_k
{\partial^{n-2k}\over \partial\phi^{n-2k}}(a-\cos\tilde\theta)^{p+n-k}
\,, \label{eq:mainterm2}
\end{eqnarray}
where $[n/2]$ denotes the maximum integer not exceeding $n/2$,
and $a_k^{(n)}$ satisfies the recurrence relation,
\begin{eqnarray}
a^{(n+1)}_k &=&
{1\over p+n-k+1}a^{(n)}_k -(n-2k+2)a^{(n)}_{k-1} \,, \\
a^{(0)}_k &=& \cases{1\,, & $k=0$\,; \cr 0\,, & $k = 1,2,3,\cdots$\,. \cr}
\end{eqnarray}
This is solved to give
\begin{eqnarray}
a^{(n)}_k &=&
{\Gamma(p+1)\Gamma(n+1)\over \Gamma(p+n-k+1)\Gamma(n-2k+1)}
{(-1)^k\over 2^k k!}
\,. \label{eq:exp-a-coef}
\end{eqnarray}
Since each $\phi$-derivative is interpreted as giving one factor
of $im$ to the harmonic coefficients in Eq.~(\ref{eq:mainterm0}),
we have the mode decomposition as
\begin{eqnarray}
(\phi-\phi')^n(a-\cos\tilde\theta)^p &=&
2\pi\sum_{\ell=0}^{\infty}\sum_{m=-\ell}^{\ell}C^{(n,p)}_{\ell m}(a)
Y_{\ell m}(\theta,\phi)Y^*_{\ell m}(\theta',\phi') \,, \\
C^{(n,p)}_{\ell m}(a) &=&
\sum_{k=0}^{[n/2]} (im)^{n-2k} a^{(n)}_k C^{p+n-k}_\ell(a)
\,, \label{eq:mainterm3}
\end{eqnarray}

With the formulas (\ref{eq:Cellp}) and (\ref{eq:mainterm3}) at hand,
we can now write down the harmonic expansion
of $\tilde\xi^{2p}$
to the leading order of the local expansion around $(\theta',\phi')$.
The result is
\begin{eqnarray}
\tilde\xi^{2p}&=&
\left(a-\cos\tilde\theta+{b\over 2}(\phi-\phi')^2\right)^p
=\sum_{n=0}^{\infty} {\Gamma(p+1)\over \Gamma(p-n+1)}{1\over n!}
\left({b\over 2}\right)^n (\phi-\phi')^{2n}(a-\cos\tilde\theta)^{p-n}
\nonumber \\
&=& 2\pi\sum_{\ell=0}^{\infty}\sum_{m=-\ell}^{\ell}D^p_{\ell m}(a)
Y_{\ell m}(\theta,\phi)Y^*_{\ell m}(\theta',\phi')
\,,
\end{eqnarray}
where
\begin{eqnarray}
D^p_{\ell m}(a) &=&
\sum_{n=0}^\infty {\Gamma(p+1)\over \Gamma(p-n+1)}
{1\over n!} \left({b\over 2}\right)^n C^{(2n,p-n)}_{\ell m}(a)
\nonumber \\
&=& \sum_{n=0}^\infty \sum_{k=0}^n
{\Gamma(p+1)\Gamma(2n+1) \over \Gamma(p+n-k+1)\Gamma(2n-2k+1)}
{(-1)^k b^n \over 2^{n+k}n!k!} (-m^2)^{n-k}C^{p+n-k}_\ell(a)
\,. \label{eq:mainterm4}
\end{eqnarray}
The double sum in the last equation with respect to
$k$ and $n$ may be simplified by introducing $\bar n:=n-k$
and summing over $\bar n$ and $k$, which can be now taken
independently. Then we have
\begin{eqnarray}
D^p_{\ell m}(a) &=& \sum_{\bar n=0}^{\infty}
\left(\sum_{k=0}^\infty {\Gamma(2\bar n+2k+1)\over \Gamma(\bar n+k+1)}
{1\over k!}\left({-b\over 4}\right)^k\right)
{\Gamma(p+1) \over \Gamma(p+\bar n+1)\Gamma(2\bar n+1)}
\left({-m^2 b\over 2}\right)^{\bar n}C^{p+\bar n}_\ell(a)
\nonumber \\
&=& {1\over \sqrt{1+b}}\sum_{n=0}^{\infty} {\Gamma(p+1)\over \Gamma(p+n+1)}
{1\over n!}\left({-m^2 b\over 2(1+b)}\right)^n C^{p+n}_\ell(a)
\,. \label{eq:mainterm5}
\end{eqnarray}
Using Eq.~(\ref{eq:leading-mode0}),
the leading behavior of the mode coefficients (\ref{eq:mainterm5})
in the coincidence limit $a\to 1+0$ becomes
\begin{eqnarray}
D^p_{\ell m}(a) \to \cases{ \displaystyle
{1\over \sqrt{1+b}}{1\over -p-1}(a-1)^{p+1} \,, & if $p+1<0$;
\cr \displaystyle
{(-1)^\ell 2^{p+1} \over \sqrt{1+b}} \sum_n
{\Gamma(p+1) \Gamma(p+n+1)\over
\Gamma(p+n-\ell+1)\Gamma(p+n+\ell+2)}{1\over n!}
\left({-m^2 b\over 1+b}\right)^n  \,, & if $p+1>0$\,. \cr}
\label{eq:leading-mode1}
\end{eqnarray}
Replacing $p$ in the above by $p-1/2$, we have the formula
referred to in the text, Eq.~(\ref{eq:xipdec}).

\end{appendix}




\begin{thebibliography}{99}

\bibitem{Quinn} T. C. Quinn, Phys. Rev. {\bf D62}, 064029 (2000).
\bibitem{DeWittBrehme} B. S. DeWitt and R. W. Brehme, Ann. Phys. (N.Y.)
{\bf 9}, 220 (1960).
\bibitem{reaction} Y. Mino, M. Sasaki and T. Tanaka,
Phys. Rev. {\bf D55}, 3457 (1997),
and in Prog. Theor. Phys. Supp. {\bf 128} (1997).
\bibitem{QuinnWald} T. C. Quinn and R. M. Wald,
Phys. Rev. {\bf D60}, 064009 (1999).
\bibitem{Mano} S. Mano, H. Suzuki and E. Takasugi,
Prog. Theor. Phys. {\bf 95}, 1079 (1996).
\bibitem{MinoNakano} Y. Mino and H. Nakano,
Prog. Theore. Phys. {\bf 100}, 507 (1998).
\bibitem{Burko} L. M. Burko, Phys. Rev. Lett. {\bf 84}, 4529 (2000),
and L. Barack and L. M. Burko, Phys. Rev. {\bf D62}, 084040 (2000).
\bibitem{Lousto} C. O. Lousto, Phys. Rev. Lett. {\bf 84}, 5251 (2000).
\bibitem{Barack} L. Barack and A. Ori, Phys. Rev. {\bf D61}, 061502 (2000),
and L. Barack, Phys. Rev. {\bf D62}, 084027 (2000).
\bibitem{NakSas} H.~Nakano and M.~Sasaki, Prog. Theor. Phys. {\bf 105},
 197 (2001) [gr-qc/0010036].
\bibitem{Detweiler} S. Detweiler, gr-qc/0011039.
\bibitem{Mino} Y. Mino, talk in 3rd Capra Ranch Meeting
(June 2-9, 2000, US), \\
http://www.tapir.caltech.edu/\~{}capra3/Proceedings/mino/index.html
\bibitem{MTW} C. W. Misner, K. S. Thorne and J. A. Wheeler,
{\it Gravitation} (Freeman, San Francisco, 1973)
\bibitem{NakanoMinoSasaki} H. Nakano, Y. Mino and M. Sasaki,
Prog. Theor. Phys. {\bf 106}, 339 (2001) [gr-qc/0104012]
\bibitem{BMONS}
L. Barack, Y. Mino, H. Nakano, A. Ori and M. Sasaki, gr-qc/0111001.
\bibitem{BOnew} L. Barack and A. Ori, in preparation.
\bibitem{BarOri2} L. Barack and A. Ori, Phys. Rev. {\bf D64}, 124003 (2001).
\bibitem{HTF} A. Erdelyi et al., 
{\it Higher Transcendental Function}, (McGraw-Hill).

\end{thebibliography}
\end{document}